\documentclass[10pt]{article}  
\usepackage{graphicx}
\usepackage{amssymb}
\usepackage{amsfonts} 
\input epsf
\parskip=\medskipamount
\usepackage{color}

   \def\CaL{{\cal L}}

\def\al{\alpha}
\def\be{\beta}
\def\ga{\gamma}

\def\la{\lambda}  \def\La{\Lambda}
\def\kp{\kappa}  
\def\te{\theta}   
\def\om{\omega}   
\def\IB{\relax{\rm l\kern-.18 em B}}
\def\IC{\relax{\rm l\kern-.50 em C}}
\def\IE{\relax{\rm l\kern-.12 em E}}
\def\IK{\relax{\rm l\kern-.18 em K}}
\def\IL{\relax{\rm I\kern-.18 em L}}
\def\IN{\relax{\rm I\kern-.18 em N}}
\def\IR{\relax{\rm I\kern-.18 em R}}
\def\smallonehalf{\frac{{}_1}{{}^2}}
 
\def\\{\hfill\break}
\def\smallonehalf{\frac{{}_1}{{}^2}}
 
\font\tenfrak=eufm10  \font\sevenfrak=eufm7  \font\fivefrak=eufm5
\newfam\frakfam
\textfont\frakfam=\tenfrak
\scriptfont\frakfam=\sevenfrak
\scriptscriptfont\frakfam=\fivefrak

\catcode`@=11 \@addtoreset{equation}{section}
\newtheorem{proposition}{Proposition}

\def\wh{\widehat}
\def\wt{\widetilde}
\def\frac#1#2{{#1\over #2}}
\def\fracpd#1#2{\frac{\partial #1}{\partial #2}}
\def\ptos{\leaders\hbox to 2mm{\hfil{.}\hfil}\hfill}

\def\Hil{{\mathcal{H}}}
\def\<#1>{\langle#1\rangle}
\begin{document}

\title{   Quantization of Hamiltonian systems with a position dependent mass:  Killing  vector fields and Noether momenta approach    }

\author{ 
Jos\'e F.\ Cari\~nena$\dagger\,^{a)}$,
Manuel F.\ Ra\~nada$\dagger\,^{b)}$, and
Mariano Santander$\ddagger\,^{c)}$ \\ [2pt]
$\dagger$
   {\sl Departamento de F\'{\i}sica Te\'orica and IUMA, Facultad de Ciencias} \\
   {\sl Universidad de Zaragoza, 50009 Zaragoza, Spain}  \\   [2pt]
$\ddagger$
   {\sl Departamento de F\'{\i}sica Te\'{o}rica and IMUVa, Facultad de Ciencias} \\
   {\sl Universidad de Valladolid, 47011 Valladolid, Spain} 
} 
%%  \date{May 5, 2017}
%-------------------------------------------------------------------
%-------------------------------------------------------------------
\maketitle 
%----------------
 % \begin{quote}
 % {\tt  [Filename: \jobname.tex] }
 % \end{quote}
%----------------

\begin{abstract} 
The quantization of systems with a position dependent mass (PDM) is studied. 
We present a method that starts with the study of the existence of Killing vector fields for the PDM geodesic motion (Lagrangian with a PDM kinetic term but without any potential) and the construction of the associated Noether momenta.
Then the method considers,  as the appropriate Hilbert space, the space of  functions that are square integrable with respect to a measure related with the PDM and, after that, it establishes the quantization, not of the canonical momenta $p$, but  of the Noether momenta $P$ instead. 
The quantum Hamiltonian, that  depends on the Noether momenta,  is obtained as an Hermitian operator defined on the PDM  Hilbert space.
In the second part several systems with position-dependent mass, most of them related with nonlinear oscillators, are quantized by making use of the method  proposed in the first part. 
\end{abstract}

\begin{quote}
%----------------

{\sl Running title:}{\enskip}
Quantization, PDM, Killing vector fields  and   Noether Momenta. 

%----------------
%%  70S05   (2000-now) Lagrangian formalism and Hamiltonian formalism
%%  81Q80   (2010-now) Special quantum systems, such as solvable systems
%%  81U15   (2000-now) Exactly and quasi-solvable systems
%----------------

%%   03.65.-w	Quantum mechanics 
%%   
%%   
\end{quote}

\vfill
\footnoterule
{\noindent\small
$^{a)}${\it E-mail address:} {jfc@unizar.es } \\
$^{b)}${\it E-mail address:} {mfran@unizar.es }  \\
$^{c)}${\it E-mail address:} {mariano.santander@uva.es }
}
%---------------
\newpage
%---------------

%---------------------
%%   \tableofcontents
%---------------------

%-----------------------------------------------
%%  Section 1
\section{Introduction}
\label{section1}

Suppose we are given a one-dimensional system described, in terms of a coordinate $x$,  by a Lagrangian 
$$
  L = \frac{1}{2}\,m(x)\dot{x}^2 - V(x) \,,{\quad} x\in \mathbb{R}\,,{\quad} m(x)>0,
$$
that is, the usual constant mass $m$ is replaced by a strictly positive function of the position;
then the Hamiltonian $H$, that is given by 
$$
  H (x,p) = \frac{1}{2}\,\frac{1}{m(x)}\,p^2 + V(x)   \,, 
$$
is correctly defined. 
The point is that there is an important problem with the construction of the quantum version of $H$, 
that is, with  the transition $H  \,\to\, \wh{H}$ from the classical system to the quantum one,  
because  if the mass $m$ becomes a  function of the spatial coordinate, $m=m(x)$, then the quantum version of the mass no longer commutes with the momentum. 
Therefore, different forms of presenting the kinetic term in the Hamiltonian $H$,  as for example
$$
 T = \frac{1}{4}\,\Bigl[\frac{1}{m(x)}\,p^2 + p^2\,\frac{1}{m(x)}\Bigr]   \,,{\quad}
 T = \frac{1}{2}\,\Bigl[\frac{1}{\sqrt{m(x)}}\,p^2\,\frac{1}{\sqrt{m(x)}}\Bigr]  \,,{\quad}
 T = \frac{1}{2}\,\Bigl[p\,\frac{1}{m(x)}\,p \Bigr]\,,
$$
are equivalent at the classical level but they lead to different and nonequivalent Schr\"odinger equations.

This problem is important mainly for two reasons. 
\begin{itemize}
\item[(i)] There are a certain number of important areas, mainly related with problems on condensed-matter physics  
(electronic properties of semiconductors, liquid crystals, quantum dots, etc), in which the behaviour of the system depends of an  effective mass that is position-dependent.

\item[(ii)] From a more conceptual viewpoint, the ordering  of  factors  in the transition from a commutative to a noncommutative formalism is an old question that remains as an important open problem in the theory of quantization.
\end{itemize}

This question has been studied by many authors from different points of view that, in most of cases,
make use of the formalism $(\al, \be, \ga)$ that we present in the next paragraphs.

%-----------------------------------------------
%%  Section 1.1
\subsection{Formalism $(\al, \be, \ga)$ }\label{section11}

The formalism we call $(\al, \be, \ga)$ makes use as a starting point of a rather general form of the kinetic term that includes several possible alternatives. 
The main idea is to represent $T$ as the following expression depending on three (related) parameters
$$
 T_{\al\be\ga} = \frac{1}{4}\,\Bigl(m^\alpha\,p\, m^\beta\,p\, m^\gamma + 
 m^\gamma\,p\, m^\beta\,p\, m^\alpha\Bigr)\,,\quad \alpha+\beta+\gamma = - 1 \,. 
$$
It was introduced by von Roos in \cite{vR83}  (generalizing a previous study by BenDaniel {\sl et al} \cite{BenDaniel66})  and then used by other different authors \cite{ZhuKroe83}--\cite{RegoRodrCur16}.  
We quote the following text appearing in \cite{BagBaQuTk05}:
\begin{quote}
One of the well-known problems of the position-dependent effective mass (PDEM) Schr\"odinger equation (SE) consists on the momentum and mass-operator noncommutativity and the resultant ordering ambiguity in the kinetic energy term. 
To cope with this difficulty, it is advantageous to use the von Roos general two-parameter form of the effective mass kinetic energy operator  \cite{vR83}  which has an inbuilt Hermiticity and contains other plausible forms as special cases.   
\end{quote}

Even if not explicitly stated, it is assumed that the configuration space is $\mathbb{R}$ with its natural coordinate $x$, the Hilbert space of the quantum system is $L^2(\mathbb{R}, dx)$ and that $p$ is represented by the differential operator $p=-i\hbar \,d/dx$ in such space, and this is the reason for the symmetrization in the expression of $ T_{\al\be\ga} $.

 It is important to remark that different choices of the parameters, $\al$, $\be$, and $\ga$, (known as von Roos ambiguity parameters) lead to distinct non-equivalent quantum Hamiltonians. 
 Therefore, this formalism admits many particular cases; we first mention  that  BenDaniel {\sl et al}  \cite{BenDaniel66}  proposed (in a   study  previous to that of von Roos) $(\al=0, \be=-1, \ga=0)$; other choices  are for example,  Zhu {\sl et al}  \cite{ ZhuKroe83}  which use the choice   $(\al=-1/2, \be=0, \ga=-1/2)$,  Li {\sl et al} \cite{LiKu93}    $(\al=0, \be=\ga=-1/2)$, and Mustafa {\sl et al} \cite{MustM09}  $(\al=-1/4, \be=-1/2,\ga=-1/4)$.   
We also note that some authors simplify the number of parameters and make use of a simpler expression: For example in Ref. \cite{CruzNeN07PLa},  \cite{CruzRos09},   the expression of $T$ is 
$$
 T_{ab} = \frac{1}{2}\,\bigl(\,m^a\,p\, m^{2b}\,p\, m^a  \,\bigr) \,,\quad a+b = -1/2 \,, 
$$
while in  Ref. \cite{LevOz10} the expression of $T$ uses only a parameter $r=a$ and then $2b=-1-2r$.

L\'evy-Leblond studied this problem in \cite{LevLeb95PRa}  and, after  analyzing some different possible quantizations, he proposed  (by making use of some arguments related with the Galilei transformations) as
 the  most appropriate form for $T_{\al\be\ga}$ to carry out the quantization the choice  $(\al=0,\be=-1, \ga=0)$ that coincides with the one used in \cite{BenDaniel66}. 
 That is,  
$$
 T_{\rm{LL}} = \frac{1}{2}\,\Bigl(p\,\frac{1}{m(x)}\,p \Bigr)\,.
$$
He then asserts that if another different form of $T_{\al\be\ga}$ is chosen, then it is convenient to introduce an effective potential  $V_{\rm{eff}}(x)$ that can be obtained by addition to the potential $V(x)$ of 
an additional term $U(x)$ depending on $m(x)$  and of its derivatives $m'(x)$ and $m''(x)$ with $(\al,\be,\ga)$-dependent coefficients 
(in fact, this is a translation of the problem from the kinetic term $T$ into the potential $V$). 
As an example, if  the kinetic term $T$  is written as 
$$
 T =  \frac{1}{4}\,\Bigl(\frac{1}{m(x)}\,p^2 + p^2\,\frac{1}{m(x)}\Bigr)  \,,
$$
then the Hamiltonian must be modified by replacing $V(x)$ with the following effective potential 
$$
 V_{\rm{eff}}=  V(x) - \frac{1}{2}\frac{m'^2}{m^3} + \frac{1}{4}\frac{m''}{m^2} \,. 
$$
A certain number of authors  have studied this question  and shown a certain preference 
for  L\'evy-Leblond  choice \cite{ChetDekH95}--\cite{AmirIq15a}  
(with or without the effective potential).

%--------------------------------
%%  Section 1.2
\subsection{Purpose and structure of the paper}
\label{section12} 

The aim  of this paper is to present a method of quantization of Hamiltonian systems with PDM that, although it is not totally new (in fact, it has been already applied in some very particular cases \cite{CRS04Rmp, CRS07AnnPhys1}), it is now presented in a general form. 
It is formulated  starting with two important points. 
First, we consider that if the constant mass $m$ is replaced by a positive function then it is convenient to introduce this function as a factor on the metric and this property has an important influence in the form of the Hilbert space of wave functions, and second,
we consider that for obtaining the quantum Hamiltonian a previous step is the quantization of the Noether momenta.

 The structure of the paper is as follows.   
  In the next Section 2 we present the main characteristics of the method of quantization of Hamiltonian systems with a PDM by making use of Killing vector fields  and Noether momenta. 
 The rest of the paper is devoted to illustrate this method with some different particular systems.
In Section  3 a nonlinear  oscillator with quasi-harmonic behaviour is studied. 
In Section 4 we study the quantization of three nonlinear oscillators with a position-dependent mass 
and in Section 5 we consider the relation with the Laplace-Beltrami quantization formalism.
We conclude in the last section with some remarks and open problems.

%-----------------------------------------------
%%  Section 2
\section{Quantization by making use of Killing vector fields and Noether momenta}
\label{section2} 

%-----------------------------------------------
%%  Section 2.1
\subsection{Killing vector fields and Noether momenta } 
\label{section21} 

In order to study a quantum system (in the  Schr\"odinger picture) we should first fix the Hilbert space $\Hil$ and then the (essentially) selfadjoint 
 operators corresponding to the relevant observables to be quantized. Recall that there are obstructions for the quantization of all classical observables (see e.g. \cite{G46}),
 and sometimes  we are only interested in the explicit form of the Hamiltonian quantum operator.
 
The domain of quantum  operators is quite important in the non-bounded case. So,  the selfadjoint  character of an operator depends not only of the formal appearance of the operator, but also of the particular 
domain of the Hilbert space in which it is defined.  The same formal aspect of  an operator can lead to a selfadjoint operator in a case, or not selfadjoint in the other. 
Therefore the quantization of the  Hamiltonian of a system means two items:
\begin{itemize}
\item[(a)] Definition of the appropriate Hilbert space of pure states.  
\item[(b)] Construction of the quantum  Hamiltonian 
(defined in the Hilbert space (a)). 
\end{itemize}

Unfortunately many authors  go directly to the point (b) without a previous detailed analysis of the point (a). 
In the problem we are going to consider (quantization of a Hamiltonian system with a PDM)  the point (a) is of a great  importance because the particular form of the
  measure $d\mu$ defining the Hilbert space $L^2(\mathbb{R}, d\mu)$   strongly depends on the characteristics 
of the function $m(x)$. 

  Let us begin by considering the classical one-dimensional free-particle motion in the real line characterized by the $x$-dependent kinetic term $T$ as a Lagrangian
\begin{equation}
  L(x,v) = T(x,v) = \frac{1}{2}\,m(x)\,v^2 \,,{\quad} m(x)>0 \,, \label{Tm} 
\end{equation}
that leads to the following nonlinear differential equation
$$
 m(x)\,\ddot{x} + \frac{1}{2}\, m'(x)\,\dot{x}^2 = 0 \,,  
$$
where $ m'(x)=dm/dx$. 
As indicated in \cite{CFR} this kinetic Lagrangian possesses an exact Noether symmetry. 
In fact, the function $T$ is  not invariant under translations but under  the action of the vector field $X$ given by
\begin{equation}
 X(x) = \frac{1}{\sqrt{\,m(x)\,}}\,\,\fracpd{}{x}  \,, \label{vfX} 
\end{equation}
(displacement $\delta x=\epsilon (m(x))^{-1/2}$, in the physicists language) 
in the sense that we have
$$
 X^t\bigl(T\bigr)=0  \,,
$$
where $X^t$ denotes the tangent  lift to the velocity phase space $\mathbb{R}{\times}\mathbb{R}$ 
(that, in differential geometric terms, is the tangent bundle $TQ$ of the configuration space $Q=\mathbb{R}$)  of the vector field $X\in\mathfrak{X}(\mathbb{R})$,
$$
 X^t (x,v)=  \frac{1}{\sqrt{\,m(x)\,}}\,\Bigl(\,\fracpd{}{x}  - \Bigl(\frac{1}{2} 
  \frac{m'(x)}{m(x)}\Bigr)v\,\fracpd{}{v}\Bigr) \,.
$$

At this point we recall that given a Riemannian space $(M,g)$, with local coordinates $x^1, x^2,\dots,x^n$,  then a vector field $X$ defined on $M$ that is a symmetry of the metric  $g$ (in the sense that it satisfies ${\mathcal{L}}_{X} g = 0$ where ${\mathcal{L}}_{X}$ denotes the Lie derivative with respect to $X$)  is called  Killing vector field. 
We also recall that Killing vector fields  also preserve the volume $\Omega_g$ determined by the metric, that is, 
$$
  \Omega_g = \sqrt{|g|}\,\,dx^1 \wedge dx^2  \wedge \dots  \wedge\,dx^n \,,{\qquad} 
  { {\mathcal{L}}}_{X} \Omega_g = 0  \,, 
$$ 
where $|g|$ denotes the determinant of the matrix $g$ defining the Riemann structure.

The following proposition relates geometry with mechanics.
%-----------------------------------------------
%%  Proposition 1
\begin{proposition}
Let $(M,g)$ a Riemannian space, $X$ a vector field on $M$, 
$X^t$ the tangent lift of $X$ to $TM$, and $T_g$ the kinetic energy function defined by the metric
$$
 T_g (x,v)= {\smallonehalf} g_{ij}(x) v^i v^j \,.
$$
Then the important  property is true 
$$
  X^t(T_g) = T_{\wt{g}} \,,{\quad} \wt{g} = \CaL_{X}g \,. 
$$
\end{proposition}
For a proof of this proposition see  \cite{CGMS}.

Consequently,  $X$ is a Killing vector field for the Riemann structure $g$ if and only if 
$X^t$ is a symmetry for the associated kinetic energy function $T_g$. 
Now, we can observe that the vector  field $X$ given by (\ref{vfX}), that preserve the PDM kinetic term (\ref{Tm}),  is in fact a Killing vector field  of the one-dimensional $m$-dependent metric
$$
 g = m(x)\, dx\otimes dx   \,,{\qquad}   ds^{2} = m(x)\,dx^2 \,.
$$
  
The line element  is invariant under the flow of the vector field $X=f(x)\partial/\partial x$ when 
$$  
  f\,m'+2\, m\, f'=0\ ,
$$
and, therefore, in order to the vector field $X$ to be a Killing vector, it
should be proportional to the vector field  $X$ given by (\ref{vfX}).

The vector field $X$ represents (the infinitesimal generator of) an exact Noether symmetry for the geodesic motion. 
If we denote by $\te_L$ the Lagrangian 1-form 
$$
  \te_L  = \Bigl(\fracpd{L}{v}\Bigr)\,dx = m(x) v\,dx \,, 
$$
then the associated Noether constant of the motion $P$ for the free (geodesic) motion  is given
by
$$ P = i\bigl(X^t\bigr)\,\theta_L  = \sqrt{m(x)}\,v \,.
$$
In what follows the function $P$ will be called Noether momentum associated to the Noether symmetry determined by $X$.

%-----------------------------------------------
%%  Section 2.2
\subsection{Quasi-regular representation} 
\label{section22} 

  The Hilbert space for a quantum system with a classical configuration space $M$ is  the linear space of square integrable functions on $M$ with respect  to an appropriate measure, $L^2(M,d\mu)$. 
 In the case of a natural system the measure to be considered must be invariant under the  the Killing vector fields of the metric. 
 The reason is the following:
 
If $\Phi:G\times M\to M$ denotes the action of a Lie group $G$ on a differentiable manifold $M$, then
the associated quasi-regular representation  is given by the following action of $G$ on the set of complex functions on $M$:
$$ 
  (U(g)\psi)(x)=\psi(\Phi(g^{-1},x)).
$$
If $M$ admits an invariant measure $d\mu$ we can restrict the action on the set 
$L^2(M,d\mu)$ and the linear representation so obtained is a unitary representation,
because then
$$
\<U(g)\psi_1,U(g)\psi_2> =\int_M((U(g)\psi_1)(x))^*(U(g)\psi_2)(x)\, d\mu(x)$$
i.e. 
$$ 
\<U(g)\psi_1,U(g)\psi_2> =\int_M\,(\psi_1(\Phi(g^{-1},x)))^*\,\psi_2(\Phi(g^{-1},x))\, d\mu(x),
$$
and consequently, defining $y$ as $y=\Phi(g^{-1},x)$, we obtain 
$$
\<U(g)\psi_1,U(g)\psi_2> =\int_M(\psi_1(y))^*\,\psi_2(y)\, d\mu(\Phi(g,y)) = 
\int_M(\psi_1(y))^*\,\psi_2(y)\, d\mu(y) = \<\psi_1,\psi_2>.
$$

If a one-parameter subgroup $\gamma(t)=\exp (at)$, $a\in\mathfrak{g}$, is considered, then the fundamental vector field   $X_a\in \mathfrak{g}$,
which is given by 
$$ 
  (X\psi)(x)=\frac d{dt}\psi(\Phi(\exp(-ta),x))_{|t=0},
$$
when restricted to the subspace $L^2(M,d\mu)$ is a skew-selfadjoint  operator provided that  the measure $\mu$ is $\gamma(t)$-invariant, because $U(\gamma(t))$ is a one-parameter group of unitary transformations.
The infinitesimal generator in the regular representation is a generator for a 1-parameter group of unitary transformations, and consequently it is skew-self-adjoint operator.
Of course if we want the generators of several one-parameter groups be skew-self-adjoint, the measure defining the Hilbert space must be invariant under each 1-parameter  subgroup.

%-----------------------------------------------
%%  Section 2.3
\subsection{Quantization } 
\label{section23}

Coming back to the one-dimensional PDM system, the quantum system must be described by   the Hilbert space of square integrable functions defined in $\mathbb{R}$ 
 endowed with an invariant under $X$  measure,  $d\mu_x$,  therefore determined by the metric.
The Lebesgue measure $dx$ is not invariant under  $X=f(x)\partial/\partial x$,
 the  invariance condition for the measure $d\mu=\rho(x)\, dx$ being 
 $$   f\,\rho'+\rho f' = 0\, .
 $$

Therefore the only  measure invariant under $X$ given by (\ref{vfX}) is any multiple of 
\begin{equation}
  d\mu_x =\sqrt{m(x)}\, dx .\label{dmu}
\end{equation} 
This automatically implies that the first-order linear operator $X$ is skew-symmetric.
This means that the operator $\widehat{P}$ representing the quantum version of the 
Noether momentum $P$ must be selfadjoint, not in the standard space 
$L^2(\mathbb{R})\equiv L^2(\mathbb{R},dx)$, but in the space
 $L^2(\mathbb{R},d\mu_x)$ of  square integrable functions with respect 
 the PDM  measure $d\mu_x$.

 Using the Legendre transformation the  momentum $p$  and velocity $v$  are related  by $p = m(x)\,v$, so that the expressions of the Noether momenta and the Hamiltonian (kinetic term plus a potential)  in the phase space are  
$$
  P = \frac{1}{\sqrt{m(x)}}\,p \,, 
$$
and
$$
 H = \frac{1}{2}\,P^2 + V(x)  \,. 
$$

As we have pointed out, the generator of the infinitesimal `translation' symmetry, 
$(1/\sqrt{m(x)}) \,{d}/{dx}$, is skew-Hermitian in the space $L^2(\mathbb{R},d\mu_x)$
and therefore the transition from the classical system to the quantum one is given 
by defining the operator $\widehat{P}$ as follows 
$$
P \ \mapsto\  \widehat{P} = \frac{1}{\sqrt{m(x)}} \Bigl( -\,i\,\hbar\,\frac{d}{dx} \Bigr) \,, 
$$
so that
$$
 \frac{1}{m}\,p^2 \ \to\ -\,\hbar^2\,
 \Bigl(\frac{1}{\sqrt{m(x)}}\,\frac{d}{dx}\Bigr)
 \Bigl(\frac{1}{\sqrt{m(x)}}\,\frac{d}{dx}\Bigr) \,,
$$
in such a way that the quantum  Hamiltonian  $\widehat{H}$ is represented by the following  Hermitian (defined on the space $L^2(\mathbb{R},d\mu_x)$) operator
\begin{eqnarray*}
 \widehat{H} &=& - \frac{\hbar^2}{2}\,\Bigl(\frac{1}{\sqrt{m(x)}}\,\frac{d}{dx}\Bigr)
 \Bigl(\frac{1}{\sqrt{m(x)}}\,\frac{d}{dx}\Bigr)   +  V(x)\,,   \cr
  &=& - \frac{\hbar^2}{2}\,\frac{1}{m(x)}\,\frac{d^2}{dx^2}
  +   \frac{\hbar^2}{4}\,\Bigl(\frac{m'(x)}{m^2(x)}\Bigr)\,\frac{d}{dx}  +  V(x)\,, 
\end{eqnarray*}
and then the  Schr\"odinger equation
$ \widehat{H}\,\Psi = E\,\Psi $
becomes 
\begin{equation}
-\,\frac{\hbar^2}{2}\,\frac{1}{m(x)}\,\frac{d^2\Psi}{dx^2}
 +   \frac{\hbar^2}{4}\,\Bigl(\frac{m'(x)}{m^2(x)}\Bigr)\,\frac{d\Psi}{dx}
 + V(x)\Psi = E\,\Psi  \,.    \label{EqSch}
\end{equation}

We can summarize the method we have presented by emphasizing two important changes with respect the standard method of quantizing in the normal case of a constant mass. First,  the Hilbert space is now related with a $m(x)$-dependent measure. 
Second,  this method quantizes, not the canonical momenta $p$, but the Noether momenta $P$. 
Then the method is carried out  in two steps: 
\begin{itemize}
\item[(i)]  Study of the properties of the classical free particle with PDM (kinetic term without potential) using geometric techniques as an approach 
$$ {\rm Kinetic\ term}\ T \ \longrightarrow \ 
{\rm Killing\ vector\  field}\ X\ \longrightarrow \ {\rm Noether\ momentum}\ P
$$
\item[(ii)] Quantization (making use of the measure $d\mu_x$) first of $P$ and then of $H$
$$  {\rm Noether\ momentum}\ P \ \longrightarrow \  
{\rm Hermitian\  operator}\ \widehat{P}\ \longrightarrow \  {\rm Quantum\ Hamiltonian}\  \widehat{H} 
$$
\end{itemize}

In the following Sections, we will  illustrate this method of quantization  with a detailed  study of some particular cases.
We will focus our attention on some systems related with nonlinear versions of the harmonic oscillator.

%-----------------------------------------------
%%  Section 3
\section{Nonlinear quasi-harmonic oscillator with a PDM  }
\label{section3} 

As a first example we review the quantization of a nonlinear oscillator already studied in \cite{CRS04Rmp,CRS07AnnPhys1}.
 The nonlinear differential equation
\begin{equation}
  (1 +\la x^2)\,\ddot{x} - (\la x)\,\dot{x}^2 + \al^2\,x  = 0   \,,{\quad}  \la>0  \,,     \label{MLeq}
\end{equation}
was first studied by Mathews and Lakshmanan in \cite{MatLak74} (see also \cite{LakRaj}) as an example of a non-linear oscillator; the most remarkable property is  that its general  solution is of the form
$$
 x = A \sin(\om\,t + \phi)   \,, {\quad}
 \om^2 = \frac{\al^2}{1 + \la\,A^2}  \,,\quad\la>0\,.
$$
That is, the above equation  represents a non-linear oscillator with periodic solutions having a simple harmonic form. 
It can be proved that (\ref{MLeq})  can be obtained  from the Euler-Lagrange equation of the   Lagrangian
\begin{equation} 
  L = \frac{1}{2}\,\Bigl(\frac{1}{1 + \la\,x^2}\Bigr)\,  (\dot{x}^2 - \al^2\,x^2)  \,. \label{LML}
\end{equation} 

The generalization of this system  to $n=2$ and $n>2$ dimensions (and also for both $\la>0$ and  $\la<0$) was studied in  \cite{CaRaSS04}. Since then it has been studied by different authors
\cite{ChandraSenthil05}--\cite{Morris15QS}.

  Let us consider the following Lagrangian 
\begin{equation}
 L(x,v;\kp) = \frac{1}{2}\,\Bigl(\frac{v^2}{1 - \kp\,x^2}\Bigr) - 
\bigl(\frac{1}{2}\bigr)\,\al^2\, \Bigl(\frac{x^2}{1 - \kp\,x^2}\Bigr)  \,,
\end{equation}
where  use has been made of a change of sign, $\kp=-\lambda$ (the reason for this change is that, in the generalized higher-dimensional case, the new parameter $\kappa$ can be interpreted as a constant curvature), and the $\kp$-dependence is defined in such a way that the limit when $\kp \to 0$ is correctly defined and it  leads to the linear harmonic oscillator. 
It is a system with a PDM $m=m(x)=1/(1 - \kp\,x^2)$ that depends on the position $x$. Remark that when $\kp>0$ we must restrict the configuration space to the open interval   $(-1/\sqrt{\kp},1/\sqrt{\kp})$   in order to the mass be positive and to  avoid singularities.

 The $\kp$-dependent  kinetic term $$T_\kp(x,v)=\frac{1}{2}\,\Bigl(\frac{v^2}{1 - \kp\,x^2}\Bigr)$$ is invariant under the action of the vector field $X_\kp$ given by
$$
 X_\kp(x) = \sqrt{\,1-\kp\,x^2\,}\,\,\fracpd{}{x}  \,,
$$
in the sense that we have
$$
 X^t_\kp\Bigl(T_\kp\Bigr)=0  \,,
$$
where $X^t_\kp$ denotes the natural lift to the velocity phase space
(tangent bundle $TQ$ of the configuration space $Q$) of the
vector field $X_\kp$ in the configuration space,
$$
 X^t_\kp(x,v) = \sqrt{\,1-\kp\,x^2}\,\,\fracpd{}{x}
 - \Bigl(\frac{\kp\,x v_x}{\sqrt{1-\kp\,x^2}}\Bigr)\fracpd{}{v_x} \,.
$$
In differential geometric terms this property means that the vector field $X_\kp$ 
is a Killing vector field  of the one-dimensional metric
$$
 g = \Bigl(\frac{1}{1 - \kp\,x^2}\Bigr)\, dx\otimes dx   \,,{\qquad}  
 ds_\kp^{2} = \Bigl(\frac{1}{1 - \kp\,x^2}\Bigr)\,dx^2 \,.
$$
It  must also be considered as a Noether symmetry for the geodesic motion with an 
associated Noether constant of the motion  $P$ for the geodesic motion  that is given by 
$$
  P = i\bigl(X^t\bigr)\,\theta_L  =  \Bigl(\frac{1}{\sqrt{1-\kp\,x^2}}\Bigr)\,v   \,. 
$$

 The expression of $P$ in the phase space is
$$
 P = \sqrt{1 - \kp\,x^2}\,p_x  \,,  
$$
so that the (classical) Hamiltonian of this $\kp$-dependent oscillator can be written as follows 
$$
  H = \bigl(\frac{1}{2}\bigr)\,P^2
  + \bigl(\frac{1}{2}\bigr)\,\al^2\, \Bigl(\frac{x^2}{1 - \kp\,x^2}\Bigr)  \,. 
$$

The quantum formalism is  constructed by considering wave  functions on the real line $\mathbb{R}$ (when $\kp<0$) or in the interval $(-1/\sqrt{\kp},1/\sqrt{\kp})$, (if  $\kp>0)$,  
endowed with  the measure $d\mu_\kp$ given by
$$
  d\mu_\kp = \Bigl(\frac{1}{\sqrt{1-\kp\,x^2}}\Bigr)\,dx \,,%\label{dmu}
$$
which is the only (up to a factor) measure invariant under $X_\kp$. 
This means that the operator $\wh{P}$, representing the PDM  linear momentum, must be Hermitian  in the space $L^2(d\mu_\kp)$ of  square integrable functions with respect the PDM  measure $d\mu_\kp$ defined as 
\begin{itemize}
\item[(i)]  In the negative $\kp<0$ case, the space $L^2(d\mu_\kp)$ is $L^2(\mathbb{R},d\mu_\kp)$. 
\item[(ii)] In the positive $\kp>0$ case, the space $L^2(d\mu_\kp)$ is $L_0^2(I_\kp,d\mu_\kp)$ where $I_\kp$ denotes the interval $[-\sqrt{\kp},1/\sqrt{\kp}]$ and the subscript means that the functions must vanish at the  endpoints $x=-1/\sqrt{\kp}$ and $x=1/\sqrt{\kp}$. 
\item[(iii)] Of course in the $\kp=0$ case we recover the standard space $L^2(\mathbb{R},dx)$.  
\end{itemize}
The quantization is given by 
$$
 P\ \mapsto\ \wh{P} =\ -\,i\,\hbar\,\sqrt{1 - \kp\,x^2}\,\frac{d}{dx}  \,,
$$
so that
$$
 (1 - \kp\,x^2)\,p_x^2 \ \to\ -\,\hbar^2\,  \Bigl(\sqrt{1 - \kp\,x^2}\,\frac{d}{dx}\Bigr)
 \Bigl(\sqrt{1 - \kp\,x^2}\,\frac{d}{dx}\Bigr) \,,
$$
in such a way that the quantum version $\widehat{H}$ of the Hamiltonian
$H$ becomes
$$
 \widehat{H} = - \frac{\hbar^2}{2}\,(1 - \kp\,x^2)\,\frac{d^2}{dx^2}
 + \bigl(\frac{\hbar^2}{2}\bigr)\,\kp\,x\,\frac{d}{dx}
 + \bigl(\frac{1}{2}\bigr)\,\al^2\, \Bigl(\frac{x^2}{1 - \kp\,x^2}\Bigr) \,.
$$
Finally,  introducing dimensionless variables $(\wt{x},\wt{\kp},e)$ 
$$
 x = \Bigl(\sqrt{\frac{\hbar}{ \al}}\,\Bigr)\,\wt{x} \,,{\quad}
 \kp = \Bigl(\frac{\al}{\hbar}\Bigl)\,\wt{\kp}  \,,{\quad}
 E = (\hbar\,\al) \, e\,,  
$$
we arrive to 
\begin{itemize}
\item{}  The quantum Hamiltonian $\widehat{H}$ becomes 
\begin{equation}
\widehat{H} = \Bigl[\,
  - \frac{1}{2}\,(1 - \wt{\kp}\,\wt{x}^2)\,\frac{d^2}{d\wt{x}^2}
  + \bigl(\frac{1}{2}\bigr)\,\wt{\kp}\,\wt{x}\,\frac{d}{d\wt{x}}
  + \bigl(\frac{1}{2}\bigr)\,\Bigl(\frac{\wt{x}^2}{1 - \wt{\kp}\,\wt{x}^2}\Bigr)
  \,\Bigr] \bigl(\hbar\,\al\bigr)   \,.
\end{equation}
\item{}   The  Schr\"odinger equation reduces to the following dimensionless form 
\begin{equation}
 (1 - \wt{k}\,\wt{x}^2) \frac{d^2}{d\wt{x}^2}\,\Psi - \wt{k} \,\wt{x}\,\frac{d}{d\wt{x}}\,\Psi
 - \Bigl(\frac{\wt{x}^2}{1 - \wt{\kp}\,\wt{x}^2}\Bigr)\,\Psi + (2\,e)\,\Psi   = 0  \,.
\end{equation}

\end{itemize}

%-----------------------------------------------
%%  Section 4
\section{Three nonlinear oscillators with a position-dependent mass } 
\label{section4}

In this section we consider three particular one-dimensional nonlinear oscillators with a position dependent mass (PDM). They were studied in \cite{CruzNeN07PLa} with the formalism of creation-annihilation operators; 
now we study the quantization by making use of the method presented in Section (\ref{section2})  
as an approach.

%--------------------------------
%%  Section 4.1 
\subsection{ $\la$-dependent nonlinear oscillator no. 1}
\label{section41} 

The position dependent mass $m_1$ and the potential $V_1$ are 
$$
  m_1 = \frac{m_0}{(1 + {\lambda}^2x^2)}  {\quad}{\rm and}{\quad} 
  V_1 = \bigl(\frac{m_0}{2\lambda^2}\bigr) \,\al^2 \Bigl({\rm arcsinh}^2({\lambda}x)\Bigr)  \,,{\quad} m_0>0,  
$$
and therefore the Lagrangian is given by 
\begin{equation}
 L_1(x,v;\la) = T_{1\la}(x,v;\la)  - V_1(x;\la)  = \frac{1}{2}\,m_0\Bigl(\frac{v^2}{1 + {\lambda}^2x^2}\Bigr) - 
\bigl(\frac{m_0}{2\lambda^2}\bigr) \,\al^2 \Bigl({\rm arcsinh}^2({\lambda}x)\Bigr)  \,,
\end{equation} 
where $\lambda$ is a parameter such that $\lambda\,x$ is dimensionless. 
The $\lambda$ dependence is defined in such a way that the following limit is satisfied 
$$
  \lim_{\la \to 0}L_1(x,v;\la)  =  {\smallonehalf}\, m_0 v^2 - {\smallonehalf}\, m_0\,\al^2 x^2  \,. 
$$
The kinetic term $T_{1\la}$ is invariant under the action of the vector field $X_\la$ given by
$$
 X_\la(x) = \sqrt{\,1 + {\lambda}^2x^2\,}\,\,\fracpd{}{x}  \,,
$$
in the sense that we have
$$
 X^t_\la\Bigl(T_{1\la}\Bigr)=0  \,,
$$
where $X^t_\la$ denotes the natural lift to the velocity phase space $\mathbb{R}{\times}\mathbb{R}$ 
(tangent bundle $TQ$ of the configuration space $Q=\mathbb{R}$) of the vector field $X_\la$,
$$
 X^t_\la (x,v)= \sqrt{\,1 + {\lambda}^2x^2\,}\,\,\fracpd{}{x}
 + \Bigl(\frac{{\lambda}^2\,x\, v}{\sqrt{1 + {\lambda}^2x^2}}\Bigr)\fracpd{}{v} \,.
$$
In differential geometric terms this property means that the vector field $X_\la$ is a Killing vector field  of the one-dimensional metric
$$
 g = \Bigl(\frac{1}{1 + {\lambda}^2x^2}\Bigr)\,dx\otimes dx   \,,{\qquad}   
 ds_\la^{2} = \Bigl(\frac{1}{1 + {\lambda}^2x^2}\Bigr)\,dx^2 \,,
$$
and, from a dynamial viewpoint, $X_\la$ must be considered as a Noether symmetry for the geodesic motion. 
The associated Noether constant of the motion $P$ for such geodesic motion is given by 
$$ 
  P = i\bigl(X_\la^t\bigr)\,\theta_L  =  \Bigl(\frac{m_0}{\sqrt{1 + {\lambda}^2x^2}}\Bigr)\,v  \,. 
$$
{}From this result we obtain that the (classical) Hamiltonian of this $\la$-dependent oscillator can be written as follows 
$$
  H_1 = \bigl(\frac{1}{2\,m_0}\bigr)\,P^2
  + (\frac{m_0}{{2\la}^2})\,\al^2\,{\rm arcsinh}^2({\lambda}x)  \,,
  \quad P = \sqrt{(1 + {\lambda}^2x^2)}\,\,p  \,. 
$$

The quantum formalism is constructed with  wave functions 
defined on the real line $\mathbb{R}$ endowed with  the measure $d\mu_\la$ given by
$$
   d\mu_\la = \Bigl(\frac{1}{\sqrt{1 + {\lambda}^2x^2}}\Bigr)\,dx \,,%\label{dmu1}
$$
which is the particular measure determined by the metric and also  the only   (up to a constant factor)  measure invariant under $X_\la$. 
This means that the operator $\wh{P}$, representing the PDM  linear momentum, must be Hermitian,  not in the standard space $L^2(\mathbb{R})$, but in the space $L^2(\mathbb{R},d\mu_\la)$; 
of course   $L^2(\mathbb{R},d\mu_\la)$ reduces in the limit $\la\to 0$ to  $L^2(\mathbb{R},dx)\equiv L^2(\mathbb{R})$.
The correspondence $P\to\wh{P}$  is given by
$$
 P\ \mapsto\ \wh{P} =\ -\,i\,\hbar\,\sqrt{1 + {\lambda}^2x^2}\,\frac{d}{dx}  \,,
$$
so that
$$
 (1 + {\lambda}^2x^2)\,p^2 \ \to\ P^2\to -\,\hbar^2\,  \Bigl(\sqrt{1 + {\lambda}^2x^2}\,\frac{d}{dx}\Bigr)  \Bigl(\sqrt{1 + {\lambda}^2x^2}\,\frac{d}{dx}\Bigr) \,,  
$$
in such a way that the quantum version $\widehat{H_1}$ of the Hamiltonian $H_1$ is 
$$
 \widehat{H_1} = - \frac{\hbar^2}{2m_0}\,(1 + {\lambda}^2x^2)\,\frac{d^2}{dx^2}
 - \bigl(\frac{\hbar^2}{2m_0}\bigr)\,{\lambda}^2x\,\frac{d}{dx}
 + \bigl(\frac{m_0}{2\lambda^2}\bigr)\,\al^2 \Bigl({\rm arcsinh}^2({\lambda}x)\Bigr) \,.
$$
and then the  Schr\"odinger equation
$$ \widehat{H_1}\,\Psi = E\,\Psi \,,
$$
turns out to be 
$$
\Bigl[\,  - \frac{\hbar^2}{2m_0}\,(1 + {\lambda}^2x^2)\,\frac{d^2}{dx^2}
 - \bigl(\frac{\hbar^2}{2m_0}\bigr)\,{\lambda}^2x\,\frac{d}{dx}
 + \bigl(\frac{m_0}{2\lambda^2}\bigr) \,\al^2 \Bigl({\rm arcsinh}^2({\lambda}x)\Bigr)
  \,\Bigr]\,\Psi  = e\,(\hbar\,\al)\,\Psi  \,. 
$$
It is convenient to simplify this equation by introducing dimensionless variables $(\wt{x},\La,e)$ defined by
$$
 x = \Bigl(\sqrt{\frac{\hbar}{m_0\al}}\,\Bigr)\,\wt{x} \,,{\quad}
 \la = \Bigl(\sqrt{\frac{m_0\,\al}{\hbar}}\Bigl)\,\La  \,,{\quad}
 E =  (\hbar\,\al)\,e   \,, 
$$
in such a way that  then
\begin{itemize}
\item{}  The quantum Hamiltonian $\widehat{H_1}$ becomes 
\begin{equation}
 \widehat{H_1} = \Bigl[\,- \frac{1}{2}\,(1 + {\La}^2\wt{x}^2)\,\frac{d^2}{d\wt{x}^2}
 - \bigl(\frac{1}{2}\bigr){\La}^2\wt{x}\,\frac{d}{d\wt{x}}
 + \bigl(\frac{1}{2\La^2}\bigr)\, {\rm arcsinh}^2({\La}\wt{x}) \,\Bigr]\,(\hbar\,\al) \,.
\end{equation}
\item{}   The  Schr\"odinger equation reduces, in terms of dimensionless variables,  to the following form: 
\begin{equation}
  (1 + {\La}^2\wt{x}^2)\,\frac{d^2}{d\wt{x}^2}\,\Psi
 + {\La}^2\wt{x}\,\frac{d}{d\wt{x}}\,\Psi
 - \bigl(\frac{1}{\La^2}\bigr) \,\Bigl({\rm arcsinh}^2({\La}\wt{x})\Bigr)\,\Psi + (2\,e)\,\Psi
 = 0  \,.
\end{equation}
\end{itemize}

%--------------------------------
%%  Section 4.2
\subsection{ $\la$-dependent nonlinear oscillator no. 2 }
\label{section42} 

The position dependent mass $m_2$ and the potential $V_2$ are 
$$
  m_2 = \frac{m_0}{(1 + {\lambda}x)^2} {\quad}{\rm and}{\quad} 
 V_2 (x;\la) = \bigl(\frac{m_0}{2\lambda^2}\bigr)\,\al^2 \Bigl(\log^2(1+{\lambda}x)\Bigr)   \,, 
$$
 where we restrict the configuration space to the interval $(-1/\lambda,\infty)$. 
 Therefore the Lagrangian, that is given by 
\begin{equation}
 L_2(x,v;\la) = T_{2\la}(x,v;\la) - V_2 (x;\la) 
 = \frac{1}{2}\,m_0\Bigl(\frac{v^2}{(1 + {\lambda}x)^2}\Bigr) - 
\bigl(\frac{m_0}{2\lambda^2}\bigr)\,\al^2 \Bigl(\log^2(1+{\lambda}x)\Bigr)  \,,
\end{equation} 
is correctly defined in  $I_\la=(-1/\la,\infty)$ and also in this case the following limit is satisfied 
$$
  \lim_{\la \to 0}L_2(x,v;\la)  =  {\smallonehalf}\, m_0 v^2 - {\smallonehalf}\, m_0\,\al^2 x^2  \,. 
$$
The kinetic term $T_{2\la}$ is invariant under the action of the vector field $X_\la$ given by 
$$
 X_\la(x) = (1 + {\lambda}x)\,\fracpd{}{x}  \,,
$$
in the sense that we have
$$
 X^t_\la\Bigl(T_{2\la}\Bigr)=0  \,,
$$
where $X^t_\la$ denotes the natural lift to the velocity phase space $\mathbb{R}{\times}I_\la$ 
(tangent bundle $TQ$ of $Q=I_\la$) of the vector field $X_\la$,
$$
  X_\la^t (x,v)= (1 + {\lambda}x)\,\fracpd{}{x} + {\lambda}\, v\fracpd{}{v} \,.
$$
In differential geometric terms this property means that the vector field $X_\la$ is a 
Killing vector field  of the one-dimensional metric
$$
 g = \Bigl(\frac{1}{(1 + {\lambda}x)^2}\Bigr)\,dx\otimes dx   \,,{\qquad}   
 ds_\la^{2} = \Bigl(\frac{1}{(1 + {\lambda}x)^2}\Bigr)\,dx^2 \,.
$$
It  must also be considered as a Noether symmetry for the geodesic motion. 
The associated Noether constant of the motion $P$ for the geodesic motion is given by 
$$ 
 P = i\bigl(X_\la^t\bigr)\,\theta_L  =  \Bigl(\frac{m_0}{(1 + {\lambda}x)}\Bigr)\,v  \,, 
$$
so that the (classical) Hamiltonian of this $\la$-dependent oscillator can be written as follows 
$$
  H_2 = \bigl(\frac{1}{2m_0}\bigr)\,P^2+ \bigl(\frac{m_0}{\lambda^2}\bigr) \, \Bigl(\log^2(1+{\lambda}x)\Bigr)  \,,
  {\quad} P =  (1 + {\lambda}x)\,p  \,. 
$$

The quantum formalism is constructed with  wave functions defined on the interval 
$I_\la=(-1/\la,\infty)$ endowed with  the measure $d\mu_\la$ given by
$$
  d\mu_\la = \Bigl(\frac{1}{ (1 + {\lambda}x)}\Bigr)\,dx  \,, % \label{dmu2}
$$
which is the measure determined by the one-dimensional metric and it is the only   (up to a constant factor) measure  invariant  under $X_\la$.   
Note also that in the limit for  $\la\to 0$, $d\mu_\la$ reduces to $dx$. 
This means that the operator $\wh{P}$, representing the PDM  linear momentum, must be Hermitian,   not in the standard space $L^2(I_\la)$ with $I_\la=(-1/\la,\infty)$, but in the space $L_0^2(I_\la,d\mu_\la)$  with $d\mu_\la$ as defined above
(the subscript means that the functions must vanish at the point $x=-1/\la$).
The transition $P\to\wh{P}$  is given by
$$
 P\ \mapsto\ \wh{P} =\ -\,i\,\hbar\,(1 + {\lambda}x)\,\frac{d}{dx}  \,,
$$
so that
$$
 (1 + {\lambda}x)^2\,p^2 \ \to\ -\,\hbar^2\,  \Bigl((1 + {\lambda}x)\,\frac{d}{dx}\Bigr)
 \Bigl((1 + {\lambda}x)\,\frac{d}{dx}\Bigr) \,,
$$
in such a way that the quantum version $\widehat{H_2}$ of the Hamiltonian $H_2$ becomes
$$
 \widehat{H_2} = - \frac{\hbar^2}{2m_0}\,(1 + {\lambda}x)^2\,\frac{d^2}{dx^2}
 - \bigl(\frac{\hbar^2}{2m_0}\bigr)\,{\lambda}(1 + {\lambda}x)\,\frac{d}{dx}
 + \bigl(\frac{m_0}{2\lambda^2}\bigr)\,\al^2 \Bigl(\log^2(1+{\lambda}x)\Bigr) \,,
$$
and then the  Schr\"odinger equation
$$ \widehat{H_2}\,\Psi = E\,\Psi \,,
$$
is 
$$
\Bigl[\, - \frac{\hbar^2}{2m_0}\,(1 + {\lambda}x)^2\,\frac{d^2}{dx^2}
 - \bigl(\frac{\hbar^2}{2m_0}\bigr)\,{\lambda}(1 + {\lambda}x)\,\frac{d}{dx}
 + \bigl(\frac{m_0}{2\lambda^2}\bigr)\,\al^2 \Bigl(\log^2(1+{\lambda}x)\Bigr)
  \,\Bigr]\,\Psi  = e\,\Psi  \,,
$$
It is convenient to simplify this equation by introducing dimensionless variables $(\wt{x},\La,e)$ defined by
$$
 x = \Bigl(\sqrt{\frac{\hbar}{m_0\al}}\,\Bigr)\,\wt{x} \,,\quad
 \la = \Bigl(\sqrt{\frac{m_0\,\al}{\hbar}}\Bigl)\,\La  \,,\quad
 E =  (\hbar\,\al)\,e   \,, 
$$
in such a way that ${\lambda}x = {\Lambda}\wt{x}$ and then we see that: 
\begin{itemize}
\item{}  The quantum Hamiltonian $\widehat{H_2}$ becomes 
\begin{equation}
 \widehat{H_2} = \Bigl[\,- \frac{1}{2}\,(1 + {\La}\wt{x})^2\,\frac{d^2}{d\wt{x}^2}
 - \bigl(\frac{1}{2}\bigr) {\La}(1 + {\La}\wt{x})\,\frac{d}{d\wt{x}}
 + \bigl(\frac{1}{2\La^2}\bigr)\,\log^2(1+{\La}\wt{x} \,\Bigr]\,(\hbar\,\al) \,.
\end{equation}
\item{}   The  Schr\"odinger equation reduces, in terms of dimensionless variables,  to the following form:  
\begin{equation}
  (1 + {\La}\wt{x})^2\,\frac{d^2}{d\wt{x}^2}\,\Psi
 + {\La}(1 + {\La}\wt{x})\,\frac{d}{d\wt{x}}\,\Psi
 - \bigl(\frac{1}{\La^2}\bigr)\, \Bigl(\log^2(1+{\La}\wt{x})\Bigr)\,\Psi + (2\,e)\,\Psi
 = 0  \,.
\end{equation}
\end{itemize}

%--------------------------------
%%  Section 4.3
\subsection{ $\la$-dependent nonlinear oscillator no. 3 }
\label{section43}

The position dependent mass $m_3$ and the potential $V_3$ are 
$$
  m_3 = \frac{m_0}{(1 - {\lambda}^2x^2)^2}   {\quad}{\rm and}{\quad} 
  V_3 (x;\la)  =  \bigl(\frac{m_0}{2\lambda^2}\bigr) \,\al^2 
  \Bigl({\rm arctanh}^2({\lambda}x)\Bigr)    \,, {\quad}  \lambda\geq 0, 
$$
and therefore the Lagrangian $L_3$  is given by 
\begin{equation}
 L_3(x,v;\la) = T_{3\la}(x,v;\la) - V_3 (x;\la) 
 = \frac{1}{2}\,m_0\Bigl(\frac{v^2}{(1 - {\lambda}^2x^2)^2}\Bigr) - 
\bigl(\frac{m_0}{2\lambda^2}\bigr) \,\al^2 \Bigl({\rm arctanh}^2({\lambda}x)\Bigr)  \,,
\end{equation} 
so the dynamics is only defined in  $I_\la=(-1/\la,1/\la)$ and also in this case the  limit $\lambda\to 0$ is
$$
  \lim_{\la \to 0}L_3(x,v;\la)  =  {\smallonehalf}\, m_0 v^2 - {\smallonehalf}\, m_0\,\al^2 x^2  \,. 
$$

The kinetic term $T_{3\la}$ is invariant under the action of the vector field $X_\la$ given by
$$
  X_\la(x) =  (1 - {\lambda}^2x^2)\,\fracpd{}{x}  \,,
$$
in the sense that we have
$$
 X_\la^t\Bigl(T_{3\la}\Bigr)=0  \,,
$$
where $X^t_\la$ denotes the natural lift to the velocity phase space $\mathbb{R}{\times}I_\la$ 
(tangent bundle $TQ$ of $Q=I_\la$) of the vector field $X_\kp$,
$$
  X^t_\la (x,v) = (1 - {\lambda}^2x^2)\,\fracpd{}{x} - 2 \la^2 x v \,\fracpd{}{v} \,.
$$
In differential geometric terms this property means that the vector field $X_\la$ 
is a Killing vector field  of the one-dimensional metric
$$
 g = \Bigl(\frac{1}{(1 - {\lambda}^2x^2)^2}\Bigr)\,dx\otimes dx   \,,{\qquad}   
 ds_\la^{2} = \Bigl(\frac{1}{(1 - {\lambda}^2x^2)^2}\Bigr)\,dx^2 \,.
$$
It  must also  be considered as a Noether symmetry for the geodesic motion. 
The associated Noether constant of the motion $P$  for the geodesic motion is given by 
$$ 
   P = i\bigl(X_\la^t\bigr)\,\theta_L  =  \Bigl(\frac{m_0}{(1 - {\lambda}^2x^2)}\Bigr)\,v  \,, 
$$
and the (classical) Hamiltonian of this $\la$-dependent oscillator can be written as follows 
$$
  H_3 = \bigl(\frac{1}{2m_0}\bigr)\,P^2
  + \bigl(\frac{m_0}{2\lambda^2}\bigr) \, \Bigl({\rm arctanh}^2({\lambda}x)\Bigr)  \,,
  \quad P = (1 - {\lambda}^2x^2)\,p  \,. 
$$

The quantum formalism is constructed with  wave functions defined on the interval 
$I_\la=(-1/\la,1/\la)$ endowed with  the measure $d\mu_\la$ given by
$$
  d\mu_\la = \Bigl(\frac{1}{(1 - {\lambda}^2x^2)}\Bigr)\,dx  \,,  %\label{dmu3}
$$
which is the measure determined by the one-dimensional metric and it is the only   (up to a constant factor) measure  invariant   under $X_\la$.   
 Note also that  in the limit for  $\la\to 0$, $d\mu_\la$ reduces to $dx$. 
This means that the operator $\wh{P}$, representing the PDM  linear momentum, must be Hermitian, not in the standard space $L^2(I_\la)\equiv L^2(I_\la,dx)$ with $I_\la=(-1/\la,1/\la)$,  but in the space $L_0^2(I_\la,d\mu_\la)$ (the subscript means that the functions must vanish at the  endpoints $x=-1/\la$ and $x=1/\la$). 
The transition $P\to\wh{P}$  is given by
$$
 P\ \mapsto\ \wh{P} =\ -\,i\,\hbar\,(1 - {\lambda}^2x^2)\,\frac{d}{dx}  \,,
$$
so that
$$
 (1 + {\lambda}^2x^2)\,p^2 \ \to\ -\,\hbar^2\,  \Bigl((1 - {\lambda}^2x^2)\,\frac{d}{dx}\Bigr)\Bigl((1 - {\lambda}^2x^2)\,\frac{d}{dx}\Bigr) \,,
$$
in such a way that the quantum version $\widehat{H_3}$ of the Hamiltonian $H_3$ is 
$$
 \widehat{H_3} = - \frac{\hbar^2}{2m_0}\,(1 - {\lambda}^2x^2)^2\,\frac{d^2}{dx^2}
 + \bigl(\frac{\hbar^2}{2m_0}\bigr)\,2\la^2 x(1 - {\lambda}^2x^2)\,\frac{d}{dx}
 + \bigl(\frac{m_0}{2\lambda^2}\bigr) \,\al^2 \Bigl({\rm arctanh}^2({\lambda}x)\Bigr) \,.
$$
and then the  Schr\"odinger equation turns out to be 
$$ \widehat{H_3}\,\Psi = E\,\Psi \,,
$$
becomes 
$$
\Bigl[\,   - \frac{\hbar^2}{2m_0}\,(1 - {\lambda}^2x^2)^2\,\frac{d^2}{dx^2}
 + \bigl(\frac{\hbar^2}{2m_0}\bigr)\,2\la^2 x(1 - {\lambda}^2x^2)\,\frac{d}{dx}
 + \bigl(\frac{m_0}{2\lambda^2}\bigr) \,\al^2 \Bigl({\rm arctanh}^2({\lambda}x)\Bigr)
  \,\Bigr]\,\Psi  = E\,\Psi  \,.
$$
It is convenient to simplify this equation by introducing dimensionless variables $(\wt{x},\La,e)$ defined by
$$
 x = \Bigl(\sqrt{\frac{\hbar}{m_0\al}}\,\Bigr)\,\wt{x} \,,\quad
 \la = \Bigl(\sqrt{\frac{m_0\,\al}{\hbar}}\Bigl)\,\La  \,,\quad
 E =  (\hbar\,\al)\,e  \,,  
$$
in such a way that  then
\begin{itemize}
\item{}  The quantum Hamiltonian $\widehat{H_3}$ becomes 
\begin{equation}
 \widehat{H_3} = \Bigl[\,- \frac{1}{2}\,(1 - {\La}^2\wt{x}^2)^2\,\frac{d^2}{d\wt{x}^2} 
 - \bigl(\frac{1}{2}\bigr)2\la^2 \wt{x}(1 - {\La}^2\wt{x}^2)\,\frac{d}{d\wt{x}} 
 + \bigl(\frac{1}{2\La^2}\bigr)\, {\rm arctanh}^2({\La}\wt{x}) \,\Bigr]\,(\hbar\,\al) \,.
\end{equation}
\item{}   The  Schr\"odinger equation reduces,  in terms of dimensionless variables,  to the following form: 
\begin{equation}
  (1 - {\La}^2\wt{x}^2)^2\,\frac{d^2}{d\wt{x}^2}\,\Psi
 - 2\la^2 \wt{x}(1 - {\La}^2\wt{x}^2)\,\frac{d}{d\wt{x}}\,\Psi
 - \bigl(\frac{1}{\La^2}\bigr)  \Bigl({\rm arctanh}^2({\La}\wt{x})\Bigr) + (2\,e)\,\Psi
 = 0  \,.
\end{equation}
\end{itemize}

%-----------------------------------------------
%%  Section 5
\section{Relation with the Laplace-Beltrami quantization formalism  }
\label{section5} 

It is known  that the quantization rule $\, p_i \to \wh{p_i}\,$, with the operator $\wh{p_i}$  represented by the linear operator  
$
 \widehat{p_i} = -i\,\hbar\,({\partial}/\partial {x_i})
$
 is only correct when the configuration space $Q$ of the system is an Euclidean space (as for example  ${\mathbb{R}}^2$ or ${\mathbb{R}}^3$) and then a classical Hamiltonian $H$, assumed of mechanical  type, can be
  written as a function of the Cartesian-Rectangular coordinates $x_i$. Nevertheless, the momentum conjugate to an arbitrary generalized coordinate $q$ is, in general, not represented by $-i\,\hbar\,(\partial/\partial{q})$.   
For example, if the classical Hamiltonian $H$ is presented in spherical coordinates $(r,\te,\phi)$ then the quantization rule $p_r\to\,\widehat{p_r}$ with $\widehat{p_r}=-i\,\hbar\,(\partial/\partial{r})$ is not correct  \cite{P28}. 

In the Euclidean case, the kinetic term of the classical Hamiltonian is transformed into the Laplacian
$$
 p_x^2+p_y^2+p_z^2 \to \wh{p}_x^2 +  \wh{p}_y^2  +  \wh{p}_z^2 =   
  -\,\hbar^2\,\nabla^2    \,,{\quad}
 \nabla^2 = \fracpd{^2}{x^2} + \fracpd{^2}{y^2}  + \fracpd{^2}{z^2}  \,,  
$$
and this fact suggests that the quantification of the kinetic term of a Hamiltonian written in curvilinear (non-Cartesian) coordinates is given by the Laplace operator in such coordinates \cite{P28}.
In the non-Euclidean case, Schr\"odinger \cite{S40} and  Stevenson  \cite{S41} studied the spherical model (see also \cite{Higgs}) and Infeld and Shild \cite{IS45} studied the corresponding problem in Lobachevsky plane
(see also, e.g. \cite{BKO,KRS}), and it is now well established that we can consider an analogous rule  in Riemann spaces (see e.g. \cite{RG14} and Section 2 of \cite{SBR}).

So, if  the  configuration  space $Q$ is endowed with a Riemann metric $g$ given by 
$$
  g = g_{ij}(q)\,d q^i \otimes d q^j  \,,{\qquad}   d s^2  = g_{ij}(q) d q^i d q^j  \,,  
$$
the kinetic part of the quantum Hamiltonian is chosen to be given by 
 $$
  \wh{H}_0  =   -\,\frac{\hbar^2}{2m}\,\nabla^2    \,,  
$$    
where $\nabla^2$ denotes the Laplace-Beltrami operator 
$$
 \nabla^2 f = {\rm div}\bigl( {\rm grad}\, f\bigr)  =  
 \frac{1}{\sqrt{|g|}\,}\,\fracpd{}{q^i} \Bigl( \sqrt{|g|}\,\bigl(g^{ij}\, \fracpd{f}{q^j} \bigr)\Bigr) \,.  
$$
That is, the Laplace-Beltrami quantization formalism quantizes directly the Hamiltonian (that is, $H  \to \wh{H}$) without the previous quantization of the momenta. 
A particular example is that of the motion on a surface in $\mathbb{R}^3$, the metric been the pull-back of the Euclidean metric.

As we have stated in Section (\ref{section2}),  the kinetic term of a one-dimensional system with a PDM  can be considered as associated to a one-dimensional metric $g$ with only one component 
$$
 g_{11} = m(x)\,,\ g^{11} = \frac{1}{m(x)} \,,\   |g| = g_{11}  \,, 
$$
so that the one-dimensional version of the Laplace-Beltrami operator is given by 
$$
 \nabla^2f  = \frac{1}{\sqrt{\,m(x)\,}}\,\,\frac{d}{dx}  \Bigl[\sqrt{\,m(x)\,}   \Bigl(\frac{1}{m} \,\frac{df}{dx} \Bigl) \Bigr]   
 =   \frac{1}{m(x)}\,\frac{d^2f}{dx^2}
 -   \frac{1}{2}\,\Bigl(\frac{m'(x)}{m^2(x)}\Bigr)\,\frac{df}{dx}  \,,  
$$
and it leads to a Hamiltonian  $\wh{H}$ that  takes the form 
\begin{equation}
\wh{H}  =  -\,\frac{\hbar^2}{2}\,\frac{1}{m(x)}\,\frac{d^2}{dx^2}
 +   \frac{\hbar^2}{4}\,\Bigl(\frac{m'(x)}{m^2(x)}\Bigr)\,\frac{d}{dx}  + V(x) \,. 
\end{equation}
It is important to remark that it coincides with the Hamiltonian (\ref{EqSch}) obtained by making use of the quantization of the Noether momenta. 

It is also to be remarked that it is known since the well known paper \cite{DW} that some quantum geometry induced potential terms related to curvatures
can appear in some quantizations as constrained systems, but we can always absorbe such terms by defining a effective potential (see e.g. \cite{dC81,dC82,EE} or 
\cite{SBR} for a more recent paper).

%-----------------------------------------------------
%%  Section 6  
\section{Concluding remarks and outlook}  
\label{section6} 

We have studied the quantization of Hamiltonian systems with a position-dependent mass by making use of the quantization of the Noether momenta (instead of the canonical momenta) as an approach. 
This means a first analysis of the PDM geodesic motion (motion with PDM but without  potential) for obtaining the Noether momenta (integral of motion for the free particle but not for the total Hamiltonian). This is so because for a natural system the symplectic form only depends on the kinetic term and not of the potential term.  
In addition we have pointed out that the Hilbert space of wave functions must also depend on the PDM. 
Actually, only if an invariant measure is considered the operators obtained as generators of Killing transformations are Hermitean. 

It is important to underline that this method, that represents an approach to the problem rather different to the formalism $(\al, \be, \ga)$  (see the references mentioned in Section (\ref{section11})),  is more than a simple practical recipe;  it is in fact a method constructed on well defined  mathematical bases. 
In addition, (as was proved in Section (\ref{section5})), it leads to an expression of the quantum Hamiltonian that coincides with the one obtained by making use of the Laplace-Beltrami quantization formalism, a property which is not true anymore for the so called  $(\al, \be, \ga)$  formalism. 
Nevertheless there is an important difference:  the Laplace-Beltrami quantization formalism gives directly the expression of the Hamiltonian without the previous quantization of the momenta and the method we have studied is a two step quantization procedure (similar to the quantization of systems with constant mass): first quantization of momenta and then quantization of the Hamiltonian.

We finalize with the following comments. 
First, as we have mentioned in the Section (\ref{section1}), the interest for the Hamiltonian systems with PDM  has increased in these last years; so the method we have studied must be applied to the quantization of all these systems (we have only considered some particular examples mainly related with the harmonic oscillator). 
Second, we have limited our study to one-dimensional systems but this method must be studied in the more general case of several degrees of freedom. 
Finally, the quantization rule $\, p \to \wh{p}\,$, with the operator $\wh{p}$  represented by the linear operator   $\widehat{p} = -i\,\hbar\,{\partial}/{\partial q}$ is not valid either  on non-Euclidean spaces, or on non-Cartesian coordinates. 
How to proceed then in a non-Euclidean one-dimensional systems, where Cartesian coordinates do not exist?. 
It seems natural to consider the quantization of systems defined on curved spaces by making use of the quantization of Noether momenta.

%-----------------------------------------------------
\section*{Acknowledgments}

JFC and MFR acknowledge support from research projects MTM2015-64166-C2-1 (MINECO, Madrid)  and DGA-E24/1 (DGA, Zaragoza)  
and MS from research projects MTM2014-57129-C2-1-P  (MINECO, Madrid) and Junta de Castilla y Le\'on Project No. VA057U16.

%--------------------------------
{\small
%--------------------------------
       }
%--------------------------------------
%--------------------------------------
\end{document}